**On the origin of bulk glass forming ability in Cu-Hf, Zr alloys**

Ramir Ristić[a], Krešo Zadro[b], Damir Pajić[b], Ignacio A. Figueroa[c], Emil Babić[b]

[a] Department of Physics, University of Osijek, Trg Ljudevita Gaja 6, HR-3100 Osijek, Croatia
[b] Department of Physics, Faculty of Science, Bijenička cesta 32, HR-10002 Zagreb, Croatia
[c] Institute for Materials Research-UNAM, Ciudad Universitaria Coyoacan, C.P. 04510 Mexico, D.F., Mexico

**Abstract**
Understanding the formation of bulk metallic glasses (BMG) in metallic systems and finding a reliable criterion for selection of BMG compositions are among the most important issues in condensed matter physics and material science. Using the results of magnetic susceptibility measurements performed on both amorphous and crystallized Cu-Hf alloys (30-70 at% Cu) we find a correlation between the difference in magnetic susceptibilities of corresponding glassy and crystalline alloys and the variation in the glass forming ability (GFA) in these alloys. Since the same correlation can be inferred from data for the properties associated with the electronic structure of Cu-Zr alloys, it seems quite general and may apply to other glassy alloys based on early and late transition metals. This correlation is plausible from the free energy considerations and provides a simple way to select the compositions with high GFA.

Corresponding author: Ramir Ristić; e-mail: ramir.ristic@fizika.unios.hr

**1. Introduction**

Rapidly cooled atomic and molecular liquids can bypass crystallization and vitrify. The basic insight into the origin of this vitrification was provided by Kauzmann [1]: there is a slowdown in configurational rearrangement caused by obstruction of kinetic motion. However, in spite of a large theoretical effort (e.g. [2, 3]) there is no unifying theory of vitrification at present. As stated by Anderson [4], the nature of glass and the glass transition is probably the deepest and most interesting unsolved problem in the solid state theory (see also [5]).

While network bonding [6] in silica and chain entanglement in polymers seem plausible mechanisms that inhibit their crystallization, the corresponding mechanism for metallic systems is less clear [7]. Further, in metallic systems there is always a large contribution to the cohesive energy from the itinerant electrons, which makes their properties very sensitive to the electronic band structure. Regardless of these differences, a detailed insight into the formation of a glassy state is necessary for future understanding and applications of both insulating [8] and metallic glasses [7,9]. For the application of metallic glasses (MG) as structural and functional materials a key issue is understanding the glass forming ability (GFA), i.e. how the critical cooling rate $R_c$ (or equivalently the maximum casting thickness $d_c$) depends on components and composition of the alloy [7, 9, 10, 11]. Because of this, a search for system parameters correlating with GFA started simultaneously with the discovery of metallic glasses [12] and has accelerated upon the proliferation of bulk metallic glasses (BMG) with $d_c$~1 cm in the 1990s [7, 10, 11]. Initially, the criteria for high GFA were based on thermodynamic parameters, such as phase diagrams (e.g. deep eutectics [12]), characteristic temperatures (e.g. the reduced glass transition temperature $T_{rg}=T_g/T_l$ [13] where $T_l$ and $T_g$ are liquidus and glass transition temperature, respectively and other , similarly constructed parameters [14]) as well as the enthalpies [15, 16] and free energies/entropies [13,17]. These criteria work fairly well for some binary and ternary alloy systems (e.g. [16, 18, 19]), but perform less well for multicomponent BMGs [14]. Thus, novel criteria evoking the number of components ("confusion" principle), atomic size mismatch and packing density [10,20], volume conservation/weak chemical interactions and frustration due to competing crystalline phases [18, 21], fragility of undercooled melt [7, 22] etc. have been introduced. Furthermore, numerical simulations are widely used to associate GFA with efficient packing of atomic clusters in MGs, e.g. [23, 24], as well as to test the importance of atomic size mismatch/geometric frustration and atomic packing density in GFA [20, 25]. In spite of a big effort in developing the criteria for GFA, the discovery of BMGs with appropriate properties and values of $d_c$ is still largely a trial-and-error process.

It has been known for a long time [9, 13] that high GFA results from the combination of similar free energies of MG and the competing/primary crystallized phase(s) (CP(s)) and very different local atomic arrangements in MG and corresponding CP(s) (thus requiring extensive rearrangements of constituent atoms for the nucleation and growth of crystals, e.g. [26] without much gain in free energy). Indeed, a strong suppression of $R_c$ with decreasing free energy difference between MG and corresponding CP(s) has been found [27]. This can be visualized within the framework of potential energy landscape models (PEL, e.g. that of Debenedetti and Stillinger in [2]): similarly deep free energy minima in the glassy and corresponding crystalline states will be beneficial for amorphisation. Since at low temperatures ($T<T_g$) the free energy is dominated by the internal energy U (external pressure effects are small in solids) and U reflects the electronic band structure (EBS) in metallic systems a similar EBS in MG and the corresponding CP(s) seems to be important for high GFA [9, 18]. Thus, the comparison of the properties that are directly related to EBS (e.g. [18]) in MG and the same primary crystallized sample may reveal the GFA of a given alloy.

In what follows we report the measurements of the magnetic susceptibility ($\chi$) of glassy ($\chi_a$) and crystallized ($\chi_x$) $Hf_{100-x}Cu_x$ alloys spanning a broad composition range (30 at% ≤ x ≤ 70 at%). We find that the difference of $\Delta\chi = \chi_a - \chi_x$ is the smallest within the range of x having the highest GFA [28]. We further compare our results with the literature results for Zr-Cu alloys and find that in these alloys, in addition to $\Delta\chi$, also the difference in the linear coefficient of the low-temperature specific heat $\gamma$ (thus also the electronic density of states at the Fermi level [18]) and even the difference in resistivity show minima in the composition range that showed the highest GFA [29, 30]. Thus, the criterion linking high GFA with the similarity of the electronic structures of the glassy and crystallized state of the same alloy works for Cu-Hf, Zr alloys and may also work for other binary, ternary and multicomponent alloys between the early and late transition metals. We note that a recent [26] comparison of the short range order/local atomic structure in $Zr_{35}Cu_{65}$ and $Zr_{35}Ni_{65}$ alloys in both a glassy and a crystallized state reveals a larger difference between local atomic arrangements of glassy and crystallized $Zr_{35}Cu_{65}$ alloy, forming BMG, than that in $Zr_{35}Ni_{65}$ alloy showing an average GFA. Thus, in Zr-Cu alloys with high Cu content both conditions for high GFA [9] seem to be fulfilled.

## 2. Experimental

The $Cu_xHf_{100-x}$ (x = 30, 40, 50, 55, 60, 65, 70) glassy ribbons with similar cross-sections (~ 2.5 x 0.03 $mm^2$) and thus with the amorphous phases having broadly the same quenched-in disorder were prepared by melt-spinning fragments of arc-melted alloys in a pure He atmosphere [28]. The glassy state of as-cast ribbons was confirmed by differential scanning calorimetry (DSC) and X-ray diffraction (XRD) studies [28]. The magnetic susceptibility of glassy alloys ($\chi_a$) was measured with a Quantum Design SQUID based magnetometer in a magnetic field B ≤ 5.5T over the temperature range 5-300K [18]. The samples used for measurement of $\chi_a$ were later crystallized following the procedure similar to that previously used for crystallization of Cu-Zr glassy alloys [30]. In particular, the alloys with x ≥ 40 were heated at 10 K/min in a high purity

Ar atmosphere up to $T_a = 821$K which corresponds to the end of the first crystallization maximum in the DSC trace [28] of the alloy with x = 65 having the highest crystallization temperature ($T_x$) of all alloys. After a short dwell time (~ 5 min) at $T_a$ the samples were furnace cooled. The same annealing procedure was followed for the alloy with x = 30 having the lowest $T_x$ [18], but with $T_a = 797$K corresponding to the end of a crystallization maximum shown in the DSC trace for this alloy. Such procedures were followed in order to obtain the primary crystallized samples (e.g. [31]), i.e. to avoid the eventual transformations of primary crystallized phases and grain growth. The XRD confirmed the fully crystallized state of all samples [32] and was broadly consistent with a previous study of the crystallization of Cu-Hf alloys [33]. For Cu-Hf alloys the crystallization becomes more complex at elevated Cu contents (x ≥ 50) with simultaneous crystallization of two different crystalline phases (some of which have a complex unit cell [33]) which is similar to what observed in Cu-Zr alloys [30]. In particular, the Cu-Hf alloys with up to 40 at.% Cu showed an almost pure $Hf_2Cu$ phase, whereas that with 50 at.% Cu was a mixture of $CuHf_2$ and $Cu_{10}Hf_7$ phases The magnetic susceptibility of crystallized alloys ($\chi_x$) was measured in the same way as $\chi_a$. The measurement error was about ±2%. We note that the magnetic susceptibility and other properties of metallic glasses which are directly related to EBS are rather insensitive to the actual quenching conditions ($R_c$, thus quenched-in disorder, e.g.[18]), which is beneficial for their application as a criterion for GFA.

## 2. Results and discussion

In Fig. 1, we compare the variations with concentration of the room-temperature magnetic susceptibilities of glassy and crystallized Cu-Hf alloys. As noted earlier [17], in spite of complex composition of magnetic susceptibility in such MGs [33], a linear decrease of $\chi_a$ with x is qualitatively the same as that of $N(E_F)$ and reflects a linear decrease of the orbital paramagnetism and the Pauli paramagnetism of the d-band with Cu content.( Note that only the free-electron part of the Pauli paramagnetizm of the d-band is directly related to $N(E_F)$ [17, 33].) Apparently, a linear variation of $\chi_a$ with composition in glassy alloys does not indicate the compositions suitable for the formation of BMG in Cu-Hf alloys [17].

In striking contrast to $\chi_a$, $\chi_x$ exhibits a non-monotonous variation with x showing a maximum somewhere between x = 55 and 60. This reflects the sensitivity of the electronic band structure to crystalline structure in Cu-Hf alloys. (Qualitatively the same variations $\chi_a$ and $\chi_x$ with Cu content have been observed in Cu-Zr alloys [29].) It is more important to note that in the range of x with a high GFA in Cu-Hf alloys [15, 17, 27] ,the values of $\chi_x$ become close to those for $\chi_a$ .Thus, the difference between EBS in glassy and crystallized alloys apparently decreases. As a result ,the difference between $\chi_a$ and $\chi_x$, $\Delta\chi = |\chi_a - \chi_x|$ becomes the smallest in the range of x with the highest GFA and increases rapidly for both at higher and lower copper contents x. Indeed, as shown in Fig. 2 , the variation of $\Delta\chi$ with x in Cu-Hf alloys is nearly the same as that of the reduced glass transition temperature $T_{rg}$ (which describes quite well GFA in these alloys [17, 27]), only the maximum of $T_{rg}$/GFA corresponds to minimum of $\Delta\chi$. (A small shift between the compositions corresponding to minimum and maximum in Fig. 2 is probably due to experimental error accumulated in two magnetic susceptibility measurements, Fig. 1.) A smooth variation of data covering a very broad concentration range in Fig. 2 seems to indicate that the electronic structure effects dominate GFA in these alloys.

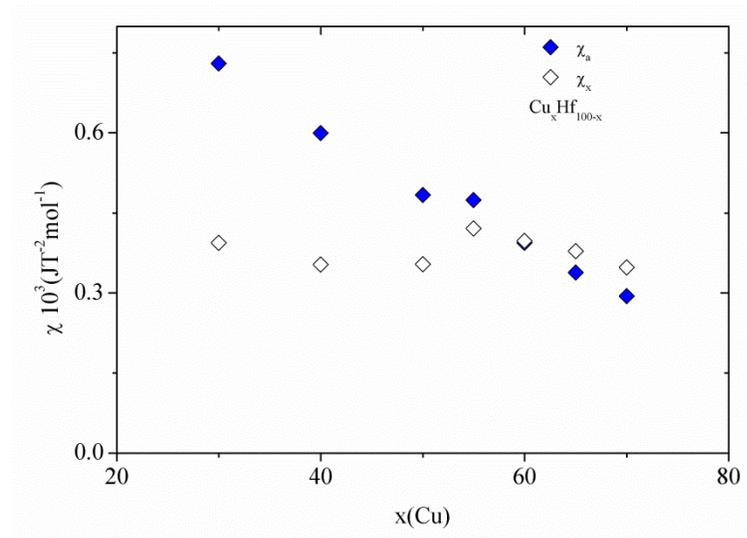

Figure 1. Magnetic susceptibilities of glassy ($\chi_a$) and crystallized ($\chi_x$) Cu-Hf alloys.

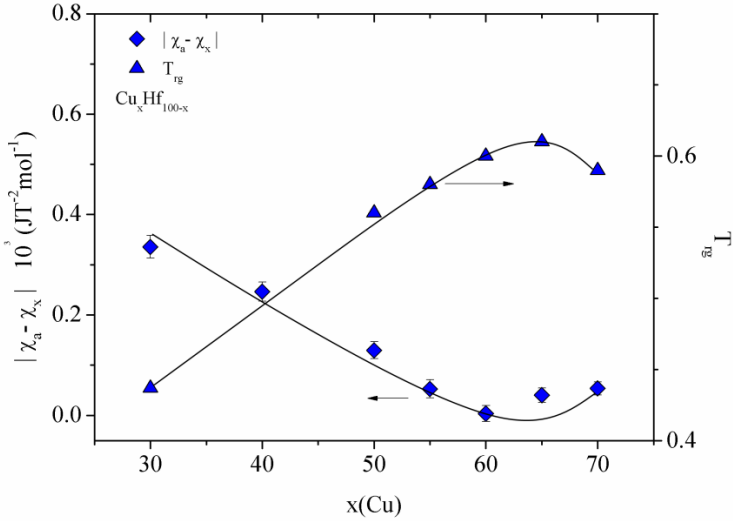

Figure 2. Change in susceptibility on crystallization, $\Delta\chi = |\chi_a - \chi_x|$, of glassy Cu-Hf alloys vs. x (left scale) and the reduced glass transition temperature, $T_{rg}=T_g/T_l$, of the same alloys vs. x (right scale).

As noted above and in the Introduction, variations of the differences between the properties related to the electronic band structure in the glassy and corresponding crystallized state observed in Cu-Zr alloys [28, 29] were very similar to those shown in Fig. 2. In particular, in addition to $\Delta\chi$, also the change in the room temperature resistivity [29], as well as that in the linear coefficient of the low temperature specific heat (LTSH), $\Delta\gamma$ (which is proportional to the difference in the dressed densities of states at the Fermi level, $\Delta N_\gamma(E_F)$), all showed minima in the concentration range with the highest GFA (thus, where BMGs can form in this alloy system). In Cu-Zr alloys also the enthalpy change in crystallization, $\Delta H_c$, was strongly reduced in the composition range showing high GFA [29], but the sensitivity of $\Delta H_c$ on eventual incipient crystallization can make the relation between measured $\Delta H_c$ and GFA unreliable.

In order to compare in a more quantitative way the results for Cu-Zr alloys [28, 29] with present results for Cu-Hf alloys, we plot in Fig. 3 the variations with composition of fractional change in $\chi$, $|\Delta\chi/\chi_a|$ and $\gamma$, $|\Delta\gamma/\gamma_a|$ for two alloy systems. We note that in the composition range with the best GFA the fractional change in $\chi$ and $\gamma$ is within about 10% and increases rapidly outside of this range. As can be expected the fractional change in susceptibility of Cu-Zr alloys shows sharp maximum at 33.3 at.% Cu where stable $CuZr_2$ compound forms directly upon crystallization [29].(Indeed, low GFA is expected at stoichiometric alloy compositions where the compound formation can occur without the necessity of a phase separation, i.e., long-range atomic rearrangements in the melt.) Further, the minima in the fractional change of $\chi$ and $\gamma$ in Cu-Zr alloys are wider than that in Cu-Hf alloys which probably reflects a broader BMG forming composition range in the former alloy system. Indeed, recent systematic research of BMG forming compositions in Cu-Zr alloys [34] indicated three compositions (x=50, 56 and 64 at.% Cu, respectively) with best GFA. ( From these alloys only that with x=50 was measured in [28, 29].) However, the variation of critical thickness within this composition range was quite small (0.6-1.2 mm) and these three small enhancements of GFA may also not be associated with EBS effects [35]. Regardless of the actual origin of these shallow maxima in GFA [34] of Cu-Zr alloys, the accumulated errors in two measurements of LTSH and magnetic susceptibility are likely to mask such a small variation of GFA in Fig.3.

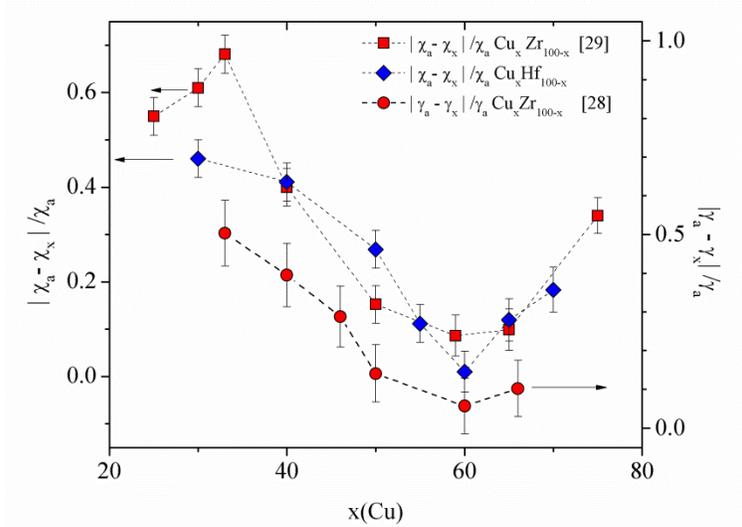

Figure 3. $|(\chi_a - \chi_x)|/\chi_a$ vs. x for Cu-Zr [29] and Cu-Hf alloys (left scale) and $|\gamma_a - \gamma_x|/\gamma_a$ vs. x for Cu-Zr [28] alloys (right scale).

The results shown in Figures 2 and 3 support a close connection between similar EBS in glassy and primary crystallized state and GFA in Cu-Zr, Hf alloys. A similar connection may exist also in other non-magnetic alloys of early transition metals with late transition metals in which the variations of properties with composition in glassy state are similar to those in Cu-Ti,

Zr, Hf alloys [36]. Further, due to some common properties of Cu-Ti, Zr, Hf glassy alloys [17, 36] and metal-metal type of multicomponent BMGs [20], the correlation between $\Delta\chi/\chi_a$ ($\Delta\gamma/\gamma_a$) and GFA may apply to these BMGs, too. Indeed , recent measurements of LTSH in multicomponent BMGs [37] indicate a small change in EBS upon primary crystallization. Further, in $Zr_{41}Ti_{14}Cu_{12.5}Ni_{10}Be_{22.5}$ BMG the difference between the density and bulk modulus in glassy and primary crystallized alloy was 1 and 3% respectively and was much smaller than that between the same parameters in the primary and equilibrium crystalline state [30].

Thus in alloy systems in which there is a substantial difference in local atomic arrangements of glassy and competing crystalline phase(s) [25], a small change in properties directly related to EBS upon crystallization may be regarded as reliable criterion for enhanced GFA and the measurements of e.g. $\chi_a$ and $\chi_x$ (or $\gamma_a$ and $\gamma_x$) may be used in order to single out the compositions with high GFA.

Since these measurements can only be made on already prepared glassy alloy it may seem that this criterion like the majority of criterions for GFA [13] serves only to speed up and simplify the exploration of the huge parameter space [38]. However, if supplemented with other criterions or research methods (e.g. [25]), this criterion may become more powerful. In particular, for binary and some ternary alloys the inspection of phase diagrams (either experimental or computed) may reveal the alloy systems and compositions with competing crystalline phases with complex unit cell and still quite low $T_l$ (thus stable liquid phase). Alternatively, the contemporary numerical simulation techniques [24, 25] may allow one to assess the atomic arrangements both in glassy and competing crystalline phase(s) as well as the total energies of these phases in a given alloy system. In these cases, simple measurements of magnetic susceptibility can be used to verify the conclusions reached by the above mentioned studies.

### 4. Conclusion

By using our results for the magnetic susceptibility of Cu-Hf alloys and the literature results for magnetic susceptibility [29], low temperature heat capacity [28] and atomic structure [25] of Cu-Zr alloys, we have shown that in both alloy systems the well known criterion for easy glass formation (similar free (internal) energies but quite different atomic arrangements in the glassy and competing crystalline state) explains quite well the variation of glass forming ability. In particular, we find that a change of magnetic susceptibility on crystallization quite accurately describes variation of glass forming ability with composition in both alloy systems. Further, we believe that this criterion may also be used in order to describe glass forming ability in non-magnetic binary and ternary alloys of Ti, Zr, Hf with late transition metals as well as in multicomponent bulk metallic glasses based on these and normal metals [37]. We also suggest that the combination of this criterion with other research methods such as the study of the phase diagrams and numerical simulations may greatly speed up and simplify the discoveries of bulk metallic glasses with desirable properties.

**Acknowledgement:** We thank Professors J.R. Cooper and B. Leontić for useful suggestions and Drs I. Bakonyi and L.K.Varga for giving us Cu-Hf samples with x=30 and 40.